\documentclass[conference]{IEEEtran}

\usepackage[utf8]{inputenc}

\usepackage{todonotes}
\usepackage{lipsum}
\usepackage{hyperref}
\usepackage{listings}
\usepackage{xcolor}
\usepackage{authblk}
\usepackage{graphicx}
\usepackage{caption}
\usepackage{subcaption}
 
\definecolor{codegreen}{rgb}{0,0.6,0}
\definecolor{codegray}{rgb}{0.5,0.5,0.5}
\definecolor{codepurple}{rgb}{0.58,0,0.82}
\definecolor{backcolour}{rgb}{0.95,0.95,0.92}

\lstdefinestyle{mystyle}{
    backgroundcolor=\color{white},   
    commentstyle=\color{codegreen},
    keywordstyle=\color{magenta},
    numberstyle=\tiny\color{codegray},
    stringstyle=\color{codepurple},
    basicstyle=\ttfamily\footnotesize,
    breakatwhitespace=false,         
    breaklines=true,                 
    captionpos=b,                    
    keepspaces=true,                 
    numbers=left,                    
    numbersep=5pt,                  
    showspaces=false,                
    showstringspaces=true,
    showtabs=false,                  
    tabsize=2
}
 
\lstdefinestyle{pythonstyle}{
    language=Python,
    basicstyle=\ttm,
    otherkeywords={self},             
    keywordstyle=\ttb\color{deepblue},
    emph={MyClass,__init__},          
    emphstyle=\ttb\color{deepred},    
    stringstyle=\color{deepgreen},
    frame=tb,                         
    showstringspaces=false            %
}

\ifCLASSINFOpdf
\else
\fi

\begin{document}
%
\title{Industrial robot ransomware: Akerbeltz}



\author[1]{Víctor Mayoral-Vilches}
\author[1]{Lander Usategui San Juan}
\author[1]{Unai Ayucar Carbajo}
\author[1]{\\Rubén Campo}
\author[1]{Xabier Sáez de Cámara}
\author[1]{Oxel Urzelai}
\author[1]{Nuria García}
\author[1]{\\Endika Gil-Uriarte}

\affil[1]{Alias Robotics, Vitoria-Gasteiz, Álava, Spain
E-mail: victor@aliasrobotics.com}

\newcommand\rednote[1]{\textcolor{red}{#1}}

\maketitle

\begin{abstract}
Cybersecurity lessons have not been learnt from the dawn of other technological industries. In robotics, the existing insecurity landscape needs to be addressed immediately. Several manufacturers profiting from the lack of general awareness are systematically ignoring their responsibilities by claiming their insecure (open) systems facilitate system integration, disregarding the safety, privacy and ethical consequences that their (lack of) actions have. In an attempt to raise awareness and illustrate the "insecurity by design in robotics" we have created  Akerbeltz, the first known instance of industrial robot ransomware. Our malware is demonstrated using a leading brand for industrial collaborative robots, Universal Robots. We describe the rationale behind our target and discuss the general flow of the attack including the initial cyber-intrusion, lateral movement and later control phase. We urge security researchers to adopt some sort of disclosure policy that forces manufacturers to react promptly. We advocate against security by obscurity and encourage the release of similar actions once vulnerability reports fall into a dead-end. Actions are now to be taken to abide a future free of zero-days for robotics.

\end{abstract}

\IEEEpeerreviewmaketitle


\section{Introduction and background}
\label{intro}

In the context of computer security, ransomware is malicious software (malware) that either locks a computer, prevents  from accessing the data using encryption, or both, until the subject has paid a ransom. First ransomware Proof of Concept (PoC) appeared in 1989 \cite{richardson2017ransomware, formby2017out} when Joseph Popp, an evolutionary biologist and AIDS researcher, carried out an experimental attack by distributing 20,000 floppy disks in a Conference by the World Health Organization focused in AIDS research. The conference spanned to researchers from more than 90 countries, and malware was distributed claiming that the disks contained a program that analyzed risk of acquiring AIDS through the use of a questionnaire. Thereafter, ransomware called "AIDS Trojan" got distributed. Since then, it has evolved leading into two big subgroups:
\begin{itemize}
    \item \textbf{Crypto ransomware}: which encrypts data and asks the user for a ransom in exchange for the decryption key.
    \item \textbf{Locker ransomware}: locks the system by some means, prevents its use and asks for a ransom to re-enable it.
\end{itemize}

According to literature \cite{richardson2017ransomware}, from 1989 to 2007 ransomware instances mostly focused on crypto-ransomware. In 2007, locker-ransomware began to appear and went mainstream. These ransomware instances locked systems and intimidated by displaying certain content (mostly pornographic images) while demanding a ransom to remove such content and unlock the systems. In 2013, Richardson et al.  \cite{richardson2017ransomware} observed that attackers pivot back to crypto-ransomware. According to the authors, the most famous piece of ransomware was released in August 2013. Named as CryptoLocker, it was originally distributed via a botnet and later through e-mail. CryptoLocker used private and public cryptographic keys to encrypt the target's file. Decrypting the files required to pay a ransom of 2 bitcoins (100 \$ at the time) within the first three days. To the extent of our literature review and to  date, variations of CryptoLocker remain being the most wide spread instances of ransomware \cite{ mcdonald2012ransomware, bhardwaj2016ransomware}.\\
\newline
In robotics, no targeted malware has yet been observed out of the PoC phase. Cesar Cerrudo and Lucas Apa earlier \cite{hackingbeforeskynet, hackingbeforeskynet2} published a ransomware attack over Nao, a social robot by Softbank Robotics, which got the media attention. According to research being conducted on the security concerns of the robotic market\footnote{\url{https://news.aliasrobotics.com/robot-security-survey-displays-first-results/} for more details on the ongoing survey.}, only 9 \% of robotics users have witnessed a cyber attack. This preliminary figure indicates that there is still very little activity yet known to the general public however, according to the same source, 51 \% of the users inquired confirmed having identified security flaws in robotic systems which leads to consider that there exists a relevant landscape of insecurities. Confirming this hypothesis, users inquired assigned a rating of 8 out 10 to the security relevance in robotics yet only 26 \% of the inquired acknowledged to have invested in robot cybersecurity, which includes evaluating security and protecting existing robot setups. Therefore, it seems that robot users do not fully apprehend the insecurity by design governing robots in the market.\\
\newline
In this paper we aim to illustrate the existing insecurity status in some robotics vendors \cite{vilches2018introducing}. We have selected one of the most popular industrial collaborative robots and present \textbf{Akerbeltz}, an instance of ransomware targeting industrial robotic systems. We present the PoC ransomware attack, describe the rationale behind our target and discuss the general flow of the attack including the initial cyber-intrusion, lateral movement and later control phase. We then briefly discuss the resulting consequences from the installation of \textbf{Akerbeltz} and wrap up by sharing some major conclusions.


\section{Akerbeltz}

In basque mythology, \emph{Akerbeltz} is an antique deity impersonated in a male goat which is the animal-kind protector. Ethimologically coming from the Basque works \emph{Aker} (male goat) and \emph{Beltz} (black), its origin is attributed to a meadow in the surroundings of Zugarramurdi caves (Navarra), a pligrimage place for Basque Mythology.  Akerbeltz is the demon that is chairing "Akelarres" or Basque witch (sorginak) meetings. Some authors note that the mythological figure Akerbeltz represents was adored in many European countries, some of which remain up to present day. Akerbeltz often shows two different faces. On the one hand, it is the protector of animals and is even able to heal their illnesses if needed. Indeed, belief on Akerbeltz is thought to be the origin of hosting a black goat as the protector of all animals within a "Baserri" (Basque cottage or farm). On the other hand, when Akerbeltz participated Akelarres, it showed his darkest face, where witches obeyed and adored him as the genius representing revolution against established status-quo, amidst banquets of human flesh.\\

We advocate for a change in (most) robot manufacturer's attitude towards security and to do so,  we take Akerbeltz as inspiration and present below the first instance of industrial robotic ransomware. Due to our concerns about malicious applications of the software, authors will not be releasing the source code nor the low level method of operation of this industrial robot ransomware. 



%

\subsection{Target selection and rationale}

\begin{figure}[h!]
    \includegraphics[width=0.45\textwidth]{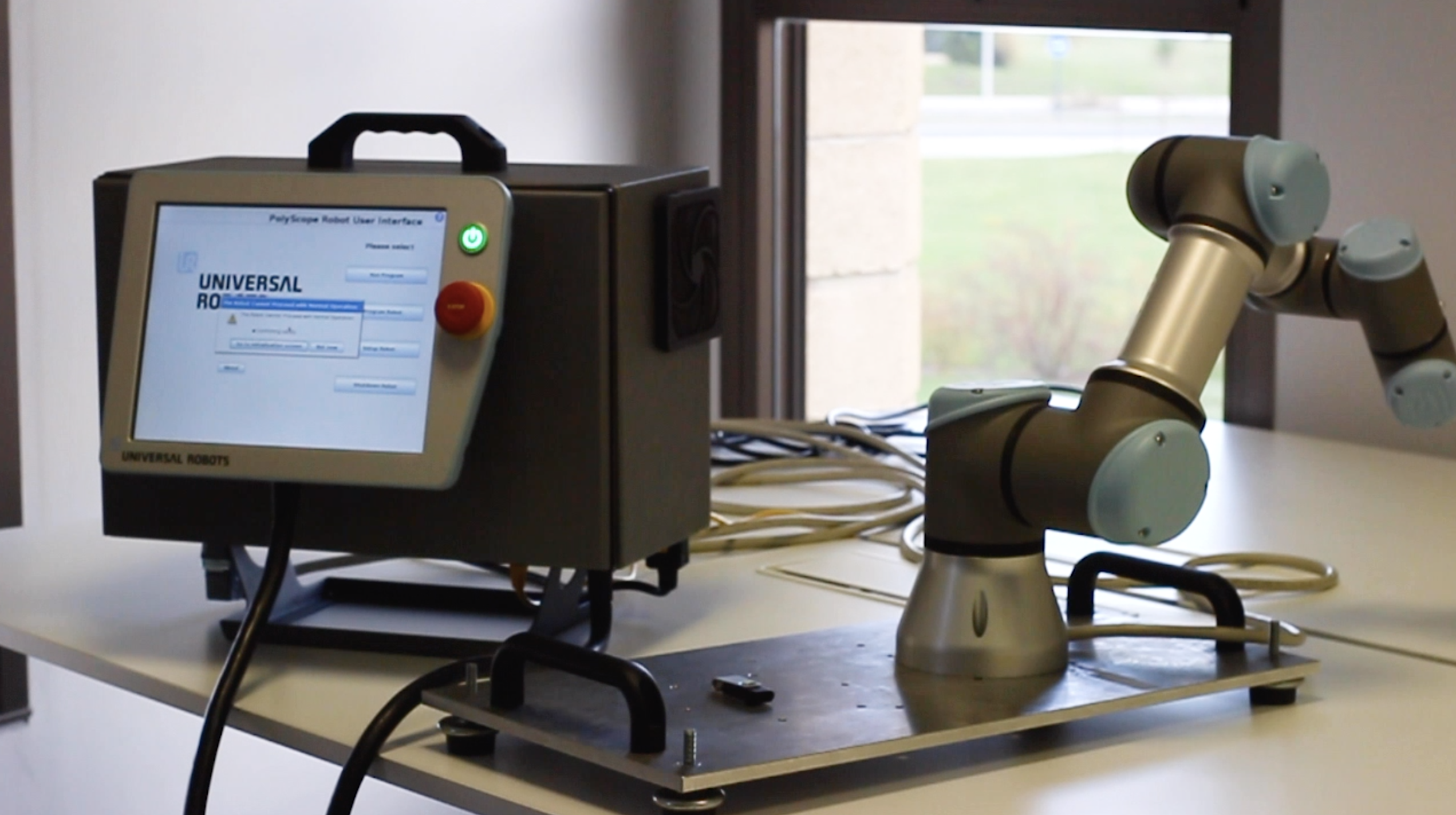}
    \centering
    \caption{Universal Robots UR3.}
    \label{fig:target}
\end{figure}

Our target was selected based on the results obtained from prior research \cite{DBLP:journals/corr/abs-1806-06681, mayoralrvd} on the vulnerability landscape across several industrial robot manufacturers. Our critical attitude was previously introduced by Alzola-Kirschgens et al. \cite{DBLP:journals/corr/abs-1806-06681} and essentially builds on the fact that several robot manufacturers, particularly collaborative robot (cobot) vendors, are profiting the popularity of these devices, via third parties (distributors and integrators) totally disregarding the ethical consequences of not caring about security, to the point that security vulnerabilities are left unadressed, or "up to the end user". To further understand our viewpoint, we ask the reader and potential robot users to consider the following questions: Thousands of insecure robots are being deployed all around the world, some of them thought to be networked and to increasingly collaborate with humans, who will be held responsible when these security holes get exploited and cause human damages? How would robot vendors respond when these reported, non-patched and public vulnerabilities cause safety hazards? Would they continue claiming that "their robots are designed to be open" and thereby never meant to be used beyond research? Should authorities regulate the use of these devices and their corresponding incomplete standards?\\
\newline
The authors discussed on all the above and coherently, decided to select what arguably is the most popular collaborative industrial robot manufacturer: Universal Robots. We select one of their best-selling robots, the UR3 and implement our PoC on it. Figure \ref{fig:target} shows a picture of our target together with the control box and teach pendant.\\

Before our work, Universal Robots had other groups assessing their insecurity. In 2017, Cerrudo and Apa reported \cite{hackingbeforeskynet, hackingbeforeskynet2} five 0-day vulnerabilities. Several months later, representatives from the vendor acknowledged that security patches had been applied \cite{robo09_2018} yet to the best of our knowledge, no public related information is available. Instead, the vendor disregarded the previously reported issues under the claim that  attackers required very specific conditions. Further to that, Jacob Bom Madsen, Software Product Manager at Universal Robots publicly claimed that Universal Robots is \emph{"proud to have a fairly open architecture, that allows system integrators and UR+ Partners to easily develop and integrate the solutions they need."}. This attitude conflicts directly with the very principle of Universal Robots safety claims, previously highlighted by Cerrudo and Apa \cite{hackingbeforeskynet2}: \emph{"Do not change anything in the safety configuration of the software (e.g. the force limit). If any safety parameter is changed the complete robot system shall be considered new, meaning that the overall safety approval process, including risk assessment, shall be updated accordingly"}. In other words, any modifications of the safety setup in the UR3 will lead to the complete invalidation of the robots' compliance with ISO 10218-1 \cite{standard10218} incurring in potential relevant losses and conflicts for the end user. Beyond the human and economical damages caused by modifying the safety setup of the UR3, the claim by Madsen leads to a troubling an arguably unethical statement: "The lack of security facilitates system integration". We have seen this statement repeatedly and yet, once again, this time coming from a leading cobot vendor, we see how openness and feature inclusion is used to justify the lack of security. Wielding the interoperability pitch, vendors push security up to their partners, collaborators or ultimately, to "the community" avoiding security actions, critical for the use of these products in human environments.\\
\newline
At the time of writing, our team knows yet of no  security patch mitigating these vulnerabilities. Moreover, we performed a penetration testing assessment in the UR3 CB series robot confirming the still unpatched existence of several of the previously reported flaws. Furthermore, we found \textbf{more than 300 new vulnerabilities} of relevant severity according to robot-specific scoring mechanisms \cite{mayoral2018towards}. The overall picture depicts a vendor which shows the little care not only for security, but also for quality of software.\\
\newline
The following subsection elaborates on how Akerbeltz acts on our UR3 CB series.

\subsection{Ransomware's flow}

\subsubsection{Cyber intrusion}

Initial infection gets realised by exploiting unpatched vulnerabilities in the robot. Akerbeltz's initial cyber intrusion is implemented exploiting these well known vulnerabilities and deployed via one of the following two attack vectors:
\begin{itemize}
    \item \textbf{physical USB ports in the teach pendant:} Exploiting \href{https://cve.mitre.org/cgi-bin/cvename.cgi?name=CVE-2019-19626}{CVE-2019-19626}\footnote{Remains confidential for responsible disclosure reasons at the time of writing} an attacker can execute malicious code with root privileges by simply connecting an external USB stick. This attack vector is likely the easiest to implement given the exposure of the teach pendant in most UR3 robots. Moreover, the same attack vector applies not only in the teach pendant but also in the control box which exposes internal USB ports that can be accessed with standard mechanical fixations. 
    
    \item \textbf{remote access via adjacent networks: } \href{https://github.com/aliasrobotics/RVD/issues/672}{RVD\#672} (\href{https://cve.mitre.org/cgi-bin/cvename.cgi?name=CVE-2018-10633}{CVE-2018-10633}) permits an adjacent network attacker to easily ssh into the control box. Alternatively,  \href{https://github.com/aliasrobotics/RVD/issues/6}{RVD\#6} allows an adjacent attacker to exploit a stack-based buffer overflow in the UR3 ModBus TCP service, and execute commands as root equally fulfilling the cyber intrusion. Akerbeltz can be configured to exploit these flaws and take action over industrial LANs.\\
\end{itemize}

\subsubsection{Lateral movement}

Since the cyber-intrusion allowed to obtain root privileges in the control box, no privilege escalation phase is necessary within Akerbeltz. The "open" architecture of the robot facilitates lateral movement to other sub-devices within the robot. We briefly explored such setup and found that it is trivial to access the robot user interface, the PolyScope. Figure \ref{fig:polyscope_hijacked} shows a preview of the ransomware's message in the teach pendant:

\begin{figure}[h!]
    \includegraphics[width=0.45\textwidth]{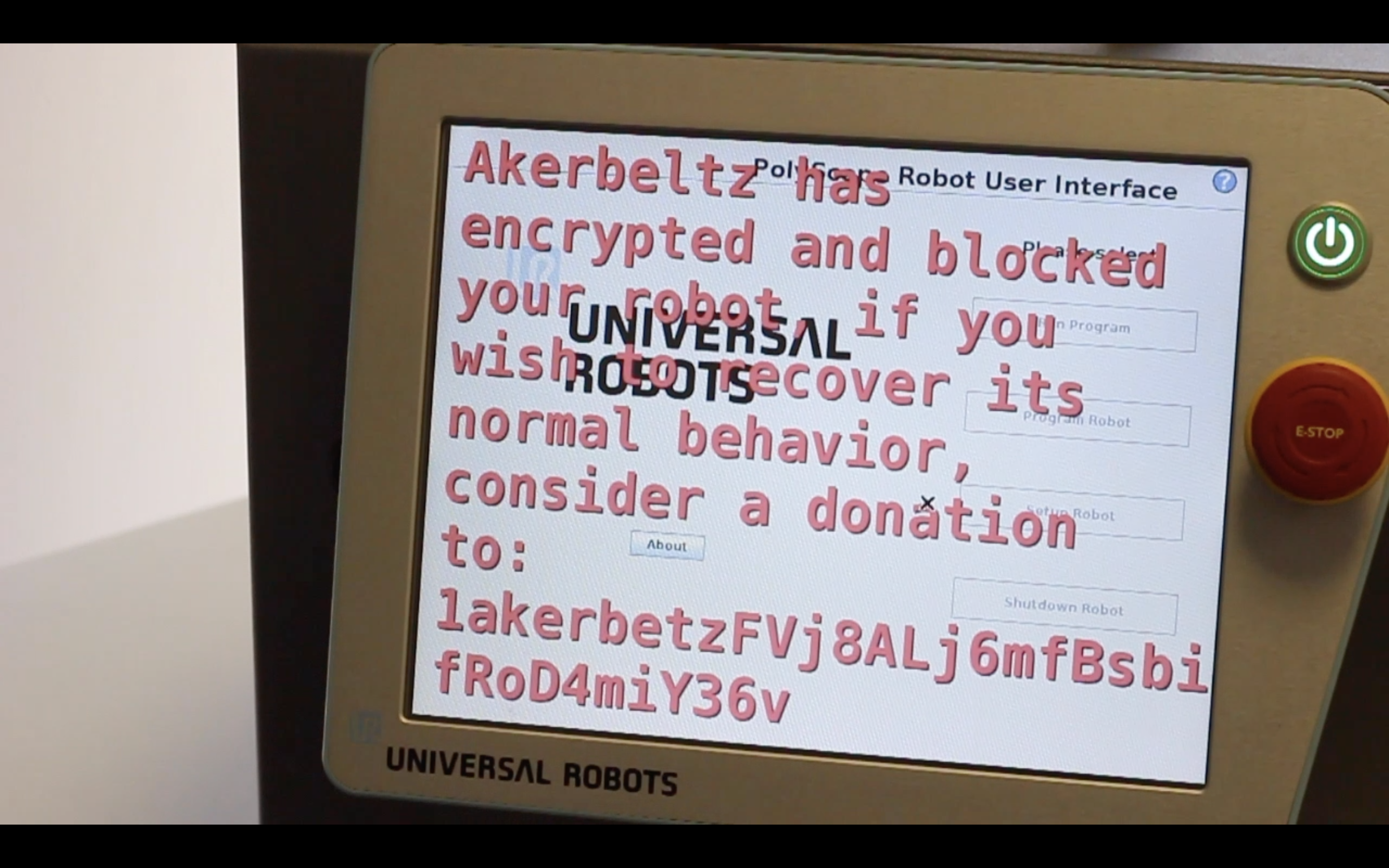}
    \centering
    \caption{UR3 PolyScope Robot User Interface  hijacked. In this case, Akerbeltz locks and disables most functions and a message is overlayed on top.}
    \label{fig:polyscope_hijacked}
\end{figure}


The UR3 lack of security allows for additional lateral movement. Both the BIOS of the robot controller and the safety PLC are easily accessible and exposed. 
While it remains beyond the scope of our study, we argue that targeting any of these systems (or both together) will likely lead to malware that could damage the robot to a point of no return and would likely be matter of future security research.\\

\subsubsection{Control}

\begin{itemize}
	\item \textbf{Locking:} In our PoC, our team was able to lock the whole system while displaying a message (refer to Figure \ref{fig:target}) after booting, using previously mentioned CVEs, we managed to access the control box, acquire root privileges, disable the default safety configuration and change the default user and password. We also identified and disabled several open ports used to control the robot from the outside, blocking other mechanisms to reestablish normal operation. Our work with Akerbeltz finalized by performing some minor and non-exhaustive hardening, meant to avoid users to unlock the system.
	\item \textbf{Encrypting:} After identifying most of the critical files used for the control of the robot, we located previous programs and IP deployed within the control box. These files are encrypted using the local pgp binary in combination with a series of robot-specific identifiers.
\end{itemize}

\subsection{Discussion}

In an attempt to responsibly disclose and mitigate the existing flaws, we generated a series of reports for selected vulnerabilities and approached the manufacturer by e-mail. To this date, no formal answer with intention to establish discussions on security has been received. In a second attempt, making use of well established and \emph{de facto} approaches in security, we filed for a CVE identifier in one of the new discovered vulnerabilities. At the time of writing, no formal communication has been established via this channel either.\\
\newline
In a third attempt, the PoC attack was disclosed to Universal Robots directly, in a public robotics industrial conference, the ROS-Industrial Conference in Europe (Stuttgart, December 2019). While maintaining the vulnerabilities undisclosed, we publicly presented\footnote{Recording of the talk is available at \url{https://youtu.be/J5-8ptUT9qU?t=27052}.} to the vendor and the rest of the audience the consequences of their insecurity. We briefly presented Akerbeltz and followed with a possible solution to mitigate existing flaws in their robots. \\
\newline
The deployment of Akerbeltz in a UR3 via the physical attack vector has been recorded and made available at \url{https://youtu.be/VF7fcV5j1t0} arguing that by making this public and available we finally call  to a reaction of this particular vendor or its associated value chain.
\newline
Last but not least, our final discussion point is that these kind of targeted attacks have the potential to spread rapidly and across the use cases in which the Universal Robots CB-series are deployed. Given the low complexity of the attack, the easiness of programming on top of Universal Robots programming environment and the relative high cost of the assets involved, a series of attacks on this taxonomy are foreseen. 

\section{Conclusions}
The current insecurity status in robotics  allows for malware to be created easily and rapidly by simply evaluating known vulnerabilities. These pieces of malware could be used and exploited by malicious actors to gain economic profits via extorsion.
In this paper we presented Akerbeltz, a piece of ransomware  that locks and encrypts industrial collaborative robots from Universal Robots. After the targeted attack, the result is rendering the industrial robot totally useless, losing IP within the robot and potentially leading to economical losses and human or environmental damages. We described and prototyped Akerbeltz for the UR3 CB series robot which could be introduced following  physical or network based attacks on zero-day vulnerabilities.\\
\newline
At the time of writing, Universal Robots has yet to answer our vulnerability reports, which in some cases provide the manufacturer relevant insights on how to mitigate the most relevant flaws exploited by the attack. As a particular example, for the physical attack vector we exploited in Akerbeltz, a possible mitigation involves the validation and sanitization of specific files as well as the implementation of Discretionary Access Control (DAC) model by providing Mandatory Access Control (MAC) within the control box. This will limit the programs' capabilities with per-program profiles (e.g. via AppArmor).\\
\newline
When asked directly, face to face, representatives of the vendor, they indicated once again that their robot "is designed to be open" or that we are "disabling features" and therefore, "it's just normal". We would like to express our strong concern for a reiterated lack of security actions. While our team did not explore malicious attacks or their profitability on the robot any further, we argue that a more aggressive individual or group, with bad intentions, might easily come up with some sort of "self-destructing" malware (by disabling safety and repeatedly crossing the boundaries of the kinematics model and hitting itself) or worse, a "human or environment damaging" one, or any other kind of creative exploits on top of the vulnerabilities exploited by Akerbeltz.\\
\newline
The lack of communication from the vendor's perspective makes us guess that, at best, there is security by obscurity around Universal Robots. Once again as we did in the past, we argue against this and advocate for a proactive interaction of robot manufacturers with security researchers.\\
\newline
Future work on our side will involve the extension of Akerbeltz and the development of RIS (Robot Immune System), an ad-hoc intrusion detection system, available for use in Universal Robots, that further monitors on top of vulnerabilities and prevents existing flaws to be exploited.

\section*{Acknowledgment}
The authors would like to thank Laura Alzola Kirschgens for her support with ethics and proofreading the content. This action was partially supported by the regional Basque Autonomous Government's SPRI agency for the support within HAZITEK funding scheme (ZL-2019/00439) and ZABALDU internationalization actions (ZAB-00014-2019). Special thanks to the Basque Cyber Security Centre BCSC for the support in actions fostering awareness in robot cyber security. Last but not least, authors are grateful to the local administration Diputación Foral de Álava for the support to entrepreneurship in innovation actions (EMPREM-2019/00002).  




%

\bibliography{bibliography}  
\bibliographystyle{IEEEtran}



\end{document}